\documentclass[amsmath,amssymb,aps,prb,twocolumn,floatfix,superscriptaddress]{revtex4-2}

\usepackage[caption = false]{subfig}
\usepackage{graphicx}
\usepackage{standalone}
\usepackage{dcolumn}
\usepackage{bm}
\usepackage{xcolor}
\usepackage{physics}
\usepackage{enumitem}
\usepackage{mathtools}
\usepackage{makecell}
\usepackage[normalem]{ulem}
\usepackage[colorlinks = true,linkcolor = red,citecolor = magenta]{hyperref}
\usepackage[sort&compress]{natbib}

\newcommand{\pcsadd}{Center for Theoretical Physics of Complex Systems, Institute for Basic Science (IBS), Daejeon 34126, Republic of Korea}
\newcommand{\ustadd}{Basic Science Program, Korea University of Science and Technology (UST), Daejeon 34113, Republic of Korea}
\newcommand{\umassadd}{Department of Physics, University of Massachusetts, 710 North Pleasant Street Amherst, MA 01003-9337, USA}

\newcommand{\mh}{\mathcal{H}}
\newcommand{\mk}{\mathcal{K}}
\newcommand{\ha}{\hat{a}}
\newcommand{\hb}{\hat{b}}
\newcommand{\hh}{\hat{h}}
\newcommand{\hv}{\hat{v}}

\newcommand{\vac}{\ket{\varnothing}}
\newcommand{\ffb}{F_\mathrm{FB}}
\newcommand{\kgsfb}{\ket{\mathrm{GS}(\ffb)}}
\newcommand{\cls}{\ket{\mathrm{CLS}}}
\newcommand{\kgs}{\ket{\mathrm{GS}}}
\newcommand{\kpsi}{\ket{\psi}}
\newcommand{\eig}{\ket{\mathrm{EIG}}}
\newcommand{\kfb}{\ket{\mathrm{FB}}}

\newcommand{\egs}{E_\mathrm{GS}}

\definecolor{myblue}{rgb}{0,0,0.75}

\begin{document}

\title{Trapping Hard-Core Bosons in Flatband Lattices}

\author{Sanghoon Lee}
    \email{scott430@naver.com}
    \affiliation{\pcsadd}
    \affiliation{\ustadd}

\author{Alexei Andreanov}
    \email{aalexei@ibs.re.kr}
    \affiliation{\pcsadd}
    \affiliation{\ustadd}

\author{Tigran Sedrakyan}
    \email{tsedrakyan@umass.edu}
    \affiliation{\umassadd}

\author{Sergej Flach}
    \email{sflach@ibs.re.kr}
    \affiliation{\pcsadd}
    \affiliation{\ustadd}

\date{\today}

\begin{abstract}
    We investigate 1D and 2D cross-stitch lattices with hard-core bosons and analytically construct exact groundstates that feature macroscopic degeneracy.
    The construction relies on the presence of a flatband in the single particle spectrum and the orthogonality of the associated compact localized states (CLS).
    Up to filling fraction \(\nu=1/2\), the groundstate is constructed by occupying the CLS.
    Exactly at \(\nu=1/2\), the groundstate becomes a Wigner crystal.
    For higher filling fractions, the groundstate is constructed by filling the CLS sites completely one by one.
    Macroscopic degeneracy arises from the multiple choices available when occupying or filling the CLS sites.
    An occupied CLS acts as an impenetrable barrier for bosons both in 1D and 2D, leading to Hilbert space fragmentation.
    A similar phenomenology also holds for hard-core bosons on the diamond chain and its higher dimensional generalizations.
    We also discuss the mapping of these hard-core models onto spin models with quantum many-body scars.
\end{abstract}

\maketitle

\section{Introduction}

Recently, systems with macroscopic degeneracies have attracted considerable interest.
This attention is due to the fragility of the degeneracies that are easily lifted even by weak perturbations, often resulting in unconventional correlated phases with interesting properties.
One particular class of such systems are flatband models~\cite{maksymenko2012flatband,leykam2018artificial,rhim2021singular}: translationally invariant tight-binding networks featuring dispersionless Bloch bands, \(E(\mathbf{k}) = E\), with zero group velocity and macroscopic degeneracy.
For short-range Hamiltonians, flatbands feature compact localized states (CLSs)---eigenstates confined to a finite number of sites~\cite{sutherland1986localization,aoki1996hofstadter}.
The CLS is a result of destructive interference caused by the network geometry, and their experimental observations have been reported in various settings~\cite{leykam2018perspective}.

The extreme sensitivity of flatbands due to their macroscopical degeneracy to perturbations and interactions gives rise to a plethora of interesting phases: flatband ferromagnetism~\cite{tasaki1992ferromagnetism}, non-conventional Landau levels~\cite{rhim2020quantum}, frustrated magnetism~\cite{ramirez1994strongly,derzhko2015strongly}, unconventional Anderson localization~\cite{goda2006inverse,flach2014detangling,longhi2021inverse}, non-perturbative metal-to-insulator~\cite{cadez2021metal,kim2023flat} or critical-to-insulator~\cite{lee2023critical,lee2023critical2,ahmed2022flat} transitions, and superconductivity~\cite{kopnin2011high,iglovikov2014superconducting,cao2018unconventional,mondaini2018pairing,aoki2020theoretical,chan2024superconductivity,chan2022pairing,chan2022designer} and quantum geometry effects~\cite{julku2021quantum,herzogarbeitman2022superfluid,tian2023evidence,torma2023essay}, among others.

Typically, adding interactions to a flatband induces transport~\cite{vidal2000interaction} via two body interaction channels, however details of the flatband, corresponding lattice symmetries~\cite{ramachandran2017chiral,calugaru2022general,regnault2022topological,bae2023isolated} and the interaction are important.
By adding a fine-tuned interaction one can observe a variety of phenomena: caging~\cite{tovmasyan2018preformed,danieli2021nonlinear,danieli2021quantum,swaminathan2023signatures}, ergodicity breaking~\cite{kuno2020flat,danieli2020many,vakulchyk2021heat}, including quantum scars~\cite{kuno2020flat_qs,hart2020compact}.
An interesting question is the fate of interacting bosons loaded in a flatband.
Additionally, Bose condensation~\cite{huber2010bose}, topological order~\cite{maiti2019fermionization}, quantum chaos and information scrambling~\cite{wei2021optical,wei2023strange} are predicted to occur in the presence of flatbands.
In three-dimensional systems, Mott insulating phase and Bose-Einstein condensation can exist~\cite{aizenman2004bose}.
However, in general, hard-core bosons can condense.
For instance, the N\'{e}el order of \(s=1/2\) magnets represents  a condensate ground state of hard-core bosons~\cite{auerbach2012interacting}.
Achieving true condensation at zero temperature is known to be unattainable for one-dimensional hard-core bosons~\cite{lenard1964momentum, lenard1966one, vaidya1979one}.
Instead, quasicondensates at finite momentum may emerge due to the presence of a quasi-long-range order in the system~\cite{rigol2004emergence, rigol2005free}.
Moreover, the violation of ETH has been observed in certain configurations of the Bose-Hubbard model in flatband lattices~\cite{danieli2020many,nicolau2023local,nicolau2023flat}.
Similarly, ETH violation has been investigated in frustrated spin systems~\cite{lee2021frustration, hahn2021information}.
Nevertheless, what happens when the groundstate of the single particle problem exhibits massive degeneracy is much less clear.

Thermalization has fascinated physicists as it describes the evolution of quantum many-body systems from reversible microscopic dynamics toward equilibrium.
Comprehensive details of this topic can be found in Ref.~\onlinecite{mori2018thermalization} and the references therein.
One intriguing aspect is the tendency of all pure states within a specific energy shell to exhibit thermal-like behavior~\cite{srednicki1994chaos}.
In search of an explanation, an eigenstate thermalization hypothesis (ETH)~\cite{deutsch2018eigenstate} has been proposed:  thermalization in isolated quantum systems can be attributed to the assumption that every eigenstate possesses thermal properties.
The concept of the ETH has received extensive attention and testing, and was confirmed in multiple settings.
Weaker forms of ETH were also proposed: weak ETH, or weak thermalization, where most but not all eigenstates exhibit thermal properties~\cite{biroli2010effect}.
It should be noted that weak thermalization alone cannot definitively establish the presence or absence of thermalization for physically realistic initial states~\cite{biroli2010effect}.
Nevertheless, weak thermalization ensures thermalisation of initial states with negligible overlap with rare athermal states.
Weak thermalization remains significant across diverse translation-invariant systems, irrespective of their integrability~\cite{alba2015eigenstate, ikeda2013finite}.
At the same time, the study of weak thermalization, including phenomena like quantum many-body scars and Hilbert space fragmentation, has been studied actively (for details, we refer to Ref.~\onlinecite{moudgalya2022quantum} and the references therein).
A unified theory of local quantum many-body dynamics was developed~\cite{doyon2017thermalization, buca2022out,buca2023unified}.

While Buca~\cite{buca2022out,buca2023unified} has established a well-constructed theoretical framework for understanding time evolution of quantum information and the spread of correlations, our study takes a distinct approach by emphasizing the role of individual eigenstates, namely CLSs in flatband systems.
Our objective is to offer a different perspective on the significance of CLSs concerning weak thermalization.
To achieve this, we delve into the thermalization characteristics of hard-core bosons within the one- and two-dimensional cross-stitch lattices.
The presence of macroscopic degeneracy in the flatband energy allows for the amplification of the effects of interactions and perturbations.
Then, our specific focus lies in the intricate interplay between CLSs and the infinite limit of strong repulsion enforced by hard-core constraints.
Our findings reveal the emergence of band-insulating and Wigner crystal phases and Hilbert space fragmentation.
The non-ergodic behaviors manifest even in the absence of disorder through CLSs, highlighting strictly non-ergodic excited states and truncation of the Hilbert space.

The paper is organized as follows.
The second section defines both one- and two-dimensional cross-stitch lattices and briefly mentions the properties of hard-core bosons.
In the third section, we explore the process of site filling to obtain the band-insulating phase and the Wigner crystal.
Moving on to the fourth and fifth sections, we investigate non-ergodic excited states in the presence of a closed CLS barrier.
Finally, conclude our work in the last section.

\section{The model}
\label{sec:model}

\begin{figure}
    \includegraphics[width = \columnwidth]{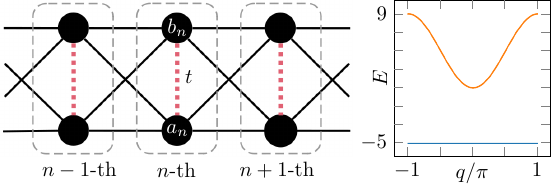}
    \caption{
        Left: Schematic of the 1D cross-stitch lattice with two sites per unit cell.
        The vertical hopping is marked by red dashed lines.
        Right: The corresponding band structure for the vertical hopping \(t = -5\).
    }
    \label{fig:1d_cs_bands}
\end{figure}

We consider hard-core bosons that obey the following set of mixed commutation relations,
\(
    [\ha_{i},\ha_{j}^{\dagger}] = \delta_{ij}(1 - 2\ha_{i}^{\dagger}\ha_{i})
\)
for all \(i\neq j\).
The commutation relations correspond to bosons with infinite onsite repulsion. 
Consequently, states with more than one particle occupying the same site are not allowed.
The hard-core bosons Hamiltonian of the one-dimensional cross-stitch chain with the nearest neighbor hopping and a vertical hopping denoted by \(t\), can be expressed as follows,
\begin{gather}
    \label{eq:cs1d}
    \mh = -\sum_{n\in\Lambda}\left[\hh_{n} + \hh_{n}^{\dagger} + t(\hv_{n} + \hv_{n}^{\dagger})\right], \\
    \hh_{n} = (\ha_{n}^{\dagger} + \hb_{n}^{\dagger})(\ha_{n+1} + \hb_{n+1}) \quad\text{and}\quad \hv_{n} = \ha_{n}^{\dagger}\hb_{n}.
\end{gather}
\(\Lambda\!\subseteq\!\mathbb{Z}\) denotes the set of unit cell indices.
The chain and the spectrum are shown in Fig.~\ref{fig:1d_cs_bands} for a specific value \(t=-5\).
In what follows, we refer to the pair of sites linked by the vertical hopping \(t\) as a dimer.
The tight-binding version of the above Hamiltonian features an orthogonal flatband whose position in the spectrum depends on the value of \(t\)~\cite{flach2014detangling}.
For \(t < -2\), the flatband becomes a groundstate.
The eigenstates of a flatband can be organized into compact localized states (CLS)~\cite{leykam2018artificial,read2017compactly}.
For an orthogonal flatband the CLSs form a complete orthonormal set.

\section{The groundstate}
\label{sec:gs}

In this section, we demonstrate the construction of the ground state for the one-dimensional cross-stitch lattice~\eqref{eq:cs1d} for arbitrary filling fractions.
We start with a generic observation: whenever the flatband is a single particle groundstate, one constructs a many body groundstate by filling non-overlapping CLS up to a critical filling fraction \(\nu_c\), where all the non-overlapping CLS are filled with one boson each.
This prescription is generic and applies to any flatband featuring a CLS~\cite{schulenburg2002macroscopic,zhitomirsky2004exact,derzhko2009exact,huber2010bose}.
However, specific models might allow for constructions extending to higher filling fractions~\cite{maiti2019fermionization}.
In the case of the cross-stitch model, we illustrate that fully-filled dimers do not contribute to the overall groundstate eigenenergy and allow for an analytical construction of a groundstate up to filling fraction \(1\).

\subsection{\texorpdfstring{\(\nu \leq 1/2\)}{nu leq 1} -- filling the non-overlapping CLS}

Consider a CLS in the one-dimensional cross-stitch lattice located at the \(m\)-th unit cell.
\begin{gather}
    \cls_{m} = \frac{\ha_{m}^{\dagger} - \hb_{m}^{\dagger}}{\sqrt{2}}\vac.
\end{gather}
\(\vac\) is a vacuum state.
Direct inspection shows that
\begin{gather}
    \label{eq:hv_cls}
    \hh_{n}\cls_{m} = \hh_{n}^{\dagger}\cls_{m} = 0, \\
    (\hv_{n} + \hv_{n}^{\dagger})\cls_{m} = -\delta_{nm}\cls_{m} \notag.
\end{gather}
We use the generic ansatz of non-overlapping CLS: since the flatband is orthogonal, we can place bosons in each unit cell independently.
We consider a set of CLS \(\ffb\) and place bosons in the respective CLS:
\begin{gather}
    \label{eq:groundstate}
    \kgsfb = \!\!\!\prod_{n\in \ffb}\!\!\!\cls_{n}.
\end{gather}
Using Eq.~\eqref{eq:hv_cls}, we see that the above is an eigenstate, and it is a groundstate since it has the same energy per particle as the single particle groundstate, the flatband:
\begin{gather}
    \label{eq:gsenergy}
    \egs = t|\ffb| = 2t\nu L,
\end{gather}
where \(\nu=N/L\) is a filling fraction which is \(\nu < 1/2\), \(N\) is the number of bosons and \(L\) is a total number of unit cells.
Moreover, thus constructed groundstates are macroscopically degenerate due to multiple ways, \(\binom{L}{N}\), to distribute \(N\) bosons over \(L\) unit cells.
For \(\nu = 1/2\), all CLS are occupied by a boson resulting in the Wigner crystal~\cite{wigner1934interaction}.

\subsection{\texorpdfstring{\(\frac{1}{2} \leq \nu \leq 1\)}{nu leq 1} -- filling the \(a_{n}, b_{n}\) dimers}

When the number of hard-core bosons exceeds the critical filling fraction \(\nu_c\), e.g., the maximum number of non-overlapping CLS giving rise to the flatband-induced Wigner crystal, the above groundstate construction fails and the groundstate can change drastically.
A single additional particle can pair up with one of the existing CLS and lead to transport, as previously discussed in Ref.~\onlinecite{drescher2017hardcore, mielke2018pair}.
However in our case, adding extra particles above the critical filling fraction \(\nu_c\) does not destroy the structure of the \(\nu\leq 1/2\) groundstate: rather CLS are gradually replaced with fully-filled dimers, thanks to the following identity for the hard-core bosons
\begin{gather}
    \ha^{\dagger}(\ha^{\dagger} - \hb^{\dagger}) \propto \ha^{\dagger}\hb^{\dagger} \propto \hb^{\dagger}(\ha^{\dagger} - \hb^{\dagger}),
\end{gather}
Therefore, we can replace some of the occupied CLSs, that denote as \(P_{d}\), with fully filled dimers.
The groundstate is given by
\begin{gather}
    \label{eq:dimerground}
    \kgs = \prod_{k\in P_{d}}\ha_{k}^{\dagger}\hb_{k}^{\dagger}\kgsfb.
\end{gather}
The size of the set \(P_{d}\) is related to the filling fraction \(\nu\) where \(1/2 < \nu \leq 1\) as follows,
\begin{gather}
    |P_{d}| = (2\nu - 1)L,
\end{gather}
where \(N\) is the total number of unit cells.
The above state is an eigenstate since the following relations hold for any \(n\)
\begin{align}
    \label{eq:barrier}
    \hh_{n}\kgs &= \hh_{n}^{\dagger}\kgs = 0, \\
    \hv_{n}\kgs &= \hv_{n}^{\dagger}\kgs = 0.
\end{align}
A fully filled dimer gives a zero contribution to the total energy.
Therefore, \(\egs\) is determined solely by the contribution of CLSs,
\begin{gather}
    \egs = t(N - |P_{d}|) = 2tN(1 - \nu).
\end{gather}
and exhibits macroscopic degeneracy due to the multiple possible choices, \(\binom{L}{2L-N}\), for positions of the filled dimers.
For \(1/2\leq \nu \leq 1\) the groundstate is a mix of fully filled dimers and the Wigner crystal.

\section{Non-ergodic excitations}
\label{sec:spinXY}

The groundstate construction outlined above also has important implications for the rest of the spectrum.
Namely, it is straightforward to check using the identites~\eqref{eq:barrier} that a boson placed next to a filled CLS cannot pass through the CLS and the latter acts as an impenetrable barrier (see Appendix~\ref{app:threebosons} for details).
This implies that a single CLS acts as an impenetrable barrier for hard-core bosons due to destructive interference.
An immediate consequence is the presence of multiple nonergodic eigenstates in the spectrum: indeed, bosons placed between a pair of filled CLS are forever trapped.
Therefore, filling a set of CLS on the rungs of the cross-stitch ladder according to some pattern would produce subspaces of Hilbert space where the memory of initial conditions is never fully lost.
Further details can be found in Appendix~\ref{app:threebosons}.
Similar results but using a different language were also proposed in Refs.~\onlinecite{buca2020quantum, buca2022out}.

\subsection{Mapping to the spin-1 XY chain}

The above cross-stitch model has a connection with the previously introduced spin-1 XY model that features quantum many-body scars~\cite{schecter2019weak}. 
To demonstrate that, we construct a mapping from hard-core bosons to spins and singlets.
We define dimer \(n\) in the original cross-stitch chain as a site of a spin chain.
For every such state (cross-stitch dimer), we define 3 triplet states \(\ket{\pm,0_t}\) corresponding to spin-1:
\begin{align}
    \ket{+}_{n} \triangleq \vert S_{n} = 1, m_{n} &= +1 \rangle = \vac, \notag \\
    \label{eq:ladder_op}
    \ket{0_{t}}_{n} \triangleq \vert S_{n} = 1, m_{n} &= \hphantom{+}0 \rangle = \frac{\ha_{n}^{\dagger} + \hb_{n}^{\dagger}}{\sqrt{2}}\vac, \\
    \ket{-}_{n} \triangleq \vert S_{n} = 1, m_{n} &= -1 \rangle = \sqrt{2}\ha^{\dagger}_{n}\hb^{\dagger}_{n} \vac, \notag
\end{align}
and 1 singlet state \(\ket{0_s}\), that corresponds to the CLS in the hard-core bosons language:
\begin{gather}
    S^{-}_{n}\cls_{n} = 0 = S^{+}_{n}\cls_{n}.
\end{gather}
Therefore, the local Hilbert space dimension of every site in the chain is 4.
The spin-1 operators are defined as follows in the hard-core bosons language:
\begin{gather}
    \label{eq:spinladder}
    S_{n}^{-} \!=\! \ha_{n}^{\dagger} \!+\! \hb_{n}^{\dagger},\hspace{0.4em} S_{n}^{+} \!=\! \ha_{n} \!+\! \hb_{n}, \hspace{0.4em}
    S_{z} \!=\! 1 \!-\! (a_{n}^{\dagger}a_{n} \!+\! b_{n}^{\dagger}b_{n}).
\end{gather}
It is straightforward to check that they satisfy the usual spin commutator algebra.
In this notation the cross-stitch Hamiltonian~\eqref{eq:cs1d} becomes the following spin-1 Hamiltonian (see Appendix~\ref{app:spinToCS}):
\begin{align}
    \label{eq:spinchain}
    \mh &= -\!\!\sum_{n\in\Lambda}S_{n}^{-}S_{n+1}^{+} \!+ S_{n+1}^{-}S_{n}^{+} \!+ t(S_{n}^{-}S_{n}^{+} \!\!+ S_{n}^{z} -1) \\
    \label{eq:spin1xy}
    &= -\!\!\sum_{\langle i,j\rangle}\!\left(S_{i}^{x}S_{j}^{x} + S_{i}^{y}S_{j}^{y}\right) + t\!\sum_{n\in\Lambda}\!\left((S_{n}^{z})^{2} - 1\right).
\end{align}
Here, we omitted the part of the Hamiltonian describing the action on the singlet states and mixed singlet-spin-1 states:
the mixed part is zero since singlets/CLS acts as barriers for hard-core bosons, while on singlet states, this Hamiltonian is diagonal and equal to \(t\), as can be verified by substituting back the definitions of the spin operators in terms of hard-core bosons.
Otherwise, this spin-1 Hamiltonian is exactly the Hamiltonian from Ref.~\onlinecite{schecter2019weak} (see the derivation in Appendix~\ref{app:spinToCS} for details.)

The results for the groundstate obtained in the previous section are easily translated into the new representation.
The triplets only state \(\ket{+ \cdots +}\) corresponds to every site being empty in the original cross-stitch chain, so that \( \mh\ket{+ \cdots +} = 0\ket{+ \cdots +} \).
Placing some singlets corresponds to filling the CLS in the bosons language. 
For the all singlets state \(\ket{0_{s} \cdots 0_{s}}\), which corresponds to the filled CLSs in every unit cell of the cross-stitch chain~\eqref{eq:cs1d}, we find Eq.~\eqref{eq:gsenergy} for \(\nu=1/2\):
\begin{gather}
    \mh\ket{0_{s} \cdots 0_{s}} = tN\ket{0_{s} \cdots 0_{s}}.
\end{gather}

The nonergodic excited states discussed for the bosons also appear naturally in the spin-singlet language.
States \(\ket{ + \cdots +  0_{s},  \Omega ,  0_{s} + \cdots + }\) form an invariant subspace of the Hilbert space under the action of \(\mh\)~\eqref{eq:spin1xy}.
We also point out that our hard-core bosons model is equivalent to the spin-1/2 XYZ Creutz ladder model studied in Ref.~\onlinecite{buca2022out}.

\begin{figure}
    \includegraphics[width = \columnwidth]{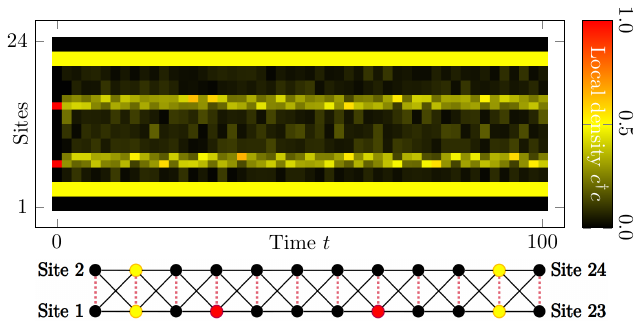}
    \caption{
        The time evolution of the initial wavefunction with four hard-core bosons on the one-dimensional cross-stitch lattice is plotted.
        Here, we set \(t = -5\).
        At time zero, two CLSs are located at the second and the 11th unit cells (yellow) with two hard-core bosons positioned at the seventh and the 15th sites (red).
        The CLSs are fixed as time evolves, and two hard-core bosons are strictly confined.
    }
    \label{fig:1d_cs_2hc}
\end{figure}

The model~\eqref{eq:spinchain} features quantum many-body scar states \(\vert\mathcal{S}_{n}\rangle\)~\cite{schecter2019weak, moudgalya2022quantum,buca2023unified}, that express \
in the hard-core boson language as 
\begin{gather}
    \vert\mathcal{S}_{n}\rangle \!\propto\!\! \sum_{\mathcal{E}_{n}}
    \left[\vphantom{\sum}(-1)^{\phi(\mathcal{E}_{n})}\!\!\prod_{m \in \mathcal{E}_{n}}\!\!\hat{a}_{m}\hat{b}_{m}\right]\prod_{j\in\Lambda}\hat{a}_{j}^{\dagger}\hat{b}_{j}^{\dagger}\vac.
\end{gather}
The \(\mathcal{E}_{n}\) is a subset of unit cell indices
\(
    \mathcal{E}_{n} = \{ j_{1}, \cdots, j_{n} \},\text{ where }j_{1}\neq\cdots\neq j_{n},
\)
and the summation runs over all the possible \(\mathcal{E}_{n}\).
The phase factor, \(\phi(\mathcal{E}_{n})\) is a sum of elements of \(\mathcal{E}_{n}\), e.g. of unit cell indices from \(\mathcal{E}_n\).
We point out that the scars \(\vert\mathcal{S}_{n}\rangle\) are unrelated to CLSs, but only involve \(\vac\) and \(\hat{a}_{n}^{\dagger}\hat{b}_{n}^{\dagger}\vac\) which correspond to the triplet states in spin language.
Moreover, it is also known to have Hilbert space fragmentation~\cite{buca2022out,moudgalya2022quantum,buca2023unified}.
CLSs play a significant role by partitioning the Hilbert space into separate Krylov subspaces, resulting in the emergence of a truncated system, as shown in Fig.~\ref{fig:1d_cs_2hc}.
However, it is important to note that the trapped bosons within the CLS barriers in Fig.~\ref{fig:1d_cs_2hc} do not constitute a quantum many-body scar because the state is a linear combination of triplet and singlet states, while the true tower of quantum many-body scars is generated from triplet states only~\cite{schecter2019weak}.
Instead, their presence gives rise to localized-like states induced by the CLS singlet state.

\section{Extension to 2D}

\begin{figure}
    \includegraphics[width = \columnwidth]{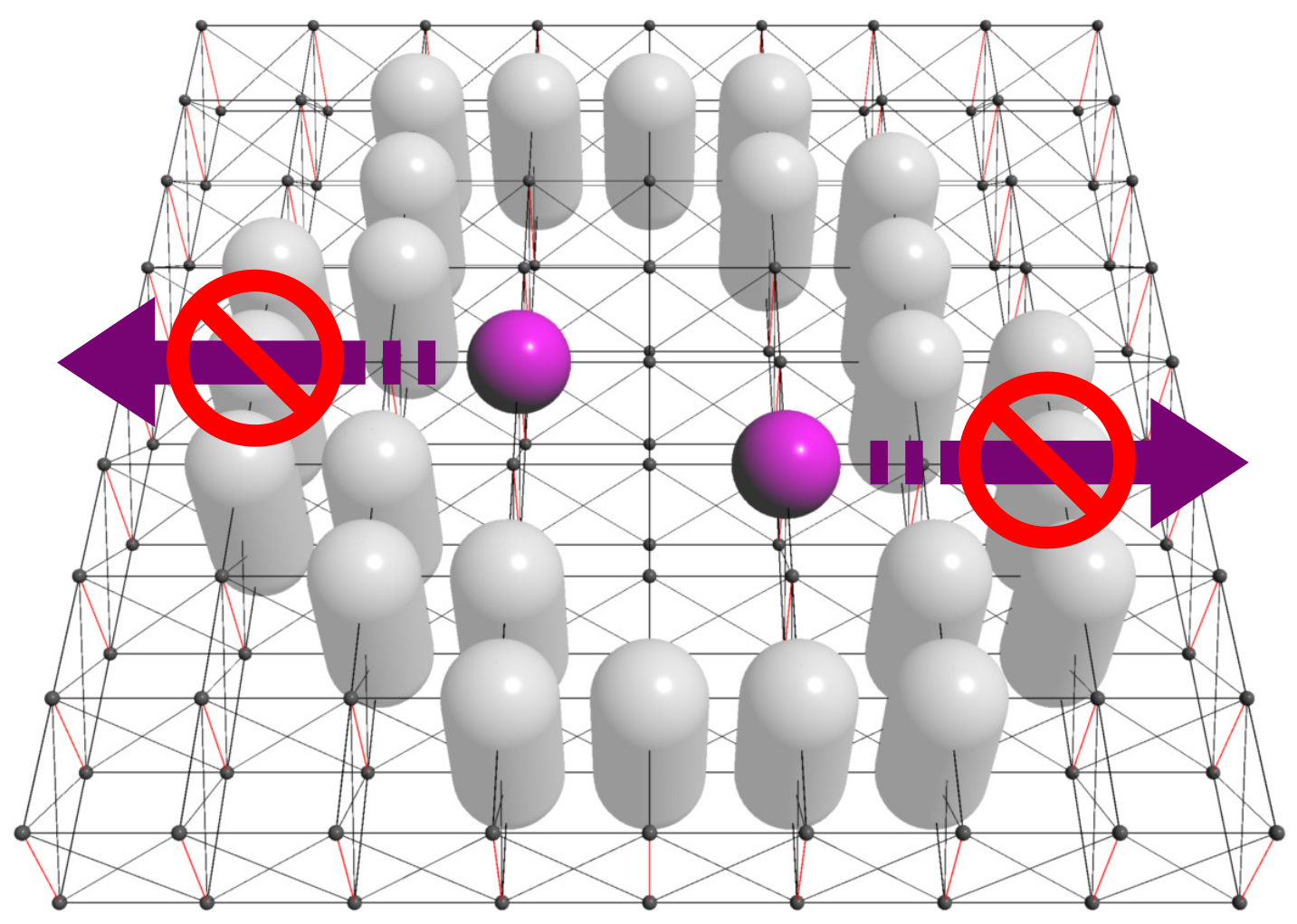}
    \caption{
        The 2D generalization of the cross-stitch chain.
        Varying the vertical hopping (red links) moves the flatband in the spectrum.
        White capsules are compact localized states of the two-dimensional cross-stitch lattice.
        Purple spheres inside the loop are hard-core bosons, each residing at specific sites.
        These bosons are restricted from moving beyond the loop, remaining constrained within the boundaries.
    }
    \label{fig:2d_cs_barrier}
\end{figure}

The results for the cross-stitch chain generalize straightforwardly to 2D, as shown in Fig.~\ref{fig:2d_cs_barrier}, which displays the 2D generalization of the cross-stitch chain.
The corresponding generalization of the 1D cross-stitch Hamiltonian for hard-core bosons with n.n. hopping reads
\begin{align}
    \label{eq:2D_ham}
    \mh & = -\sum_{n,m}\left[\hh_{n,m} + \hh_{n,m}^{\dagger} + t(\hv_{n,m} + \hv_{n,m}^{\dagger})\right], \\
    & \hh_{n,m} = (\ha_{n,m}^{\dagger} + \hb_{n,m}^{\dagger})(\ha_{n+1,m} + \hb_{n+1,m}) \notag \\
    & + (\ha_{n,m}^{\dagger} + \hb_{n,m}^{\dagger})(\ha_{n,m+1} + \hb_{n,m+1}), \notag \\
    & \hv_{n,m} = \ha_{n,m}^{\dagger}\hb_{n,m}. \notag
\end{align}
Here \(t\) is again the vertical hopping link indicated by red lines in Fig.~\ref{fig:2d_cs_barrier}.
The tight-binding version of this Hamiltonian also features an orthogonal flatband that can be freely moved around the spectrum by varying the value of \(t\).
The flatband becomes the groundstate for \(t\leq -4\).

The results for the 1D cross-stitch extend directly to this 2D model, and we only outline them here.
We consider \(N\) hard-core bosons on \(L^2\) unit cells of the 2D lattice.
The groundstate can be constructed analytically by repeating and adapting the steps of Sec.~\ref{sec:gs} to 2D.
First, for \(\nu < 1/2\), we again follow the generic prescription, and \(N\) bosons are distributed among \(L^2\) unit cells by filling CLSs.
The degeneracy of the groundstate is \(\binom{L^2}{N}\).
For half-filling \(\nu = 1/2\), the groundstate becomes the Wigner crystal.
For \( 1/2 < \nu \leq 1\), additional particles are distributed over dimers/unit-cells that become fully filled with 2 bosons.
The groundstate is a mix of fully filled dimers and the Wigner crystal.

Similarly, the filled CLS acts as a blocking barrier for other bosons inducing Hilbert space fragmentation and allowing the construction of non-ergodic eigenstates.
The corresponding derivation is a straightforward extension of the 1D cross-stitch lattice.
However, due to the 2D nature of the lattice, the resulting fragmentation is richer and more interesting.
For instance, unlike the 1D case, the barriers of filled CLS can now be organized into differently shaped loops.
Importantly, such a barrier is impenetrable for bosons even if the filled CLS are not n.n but are 2nd n.n. (on the square lattice of unit cells), e.g., placed diagonally next to each other.

It is also possible to generalize the spin-to-boson mapping using the same definitions~\eqref{eq:spinladder}.
This yields the following Hamiltonian~\cite{schecter2019weak}:
\begin{gather}
    \mh = \!\!\sum_{\langle \mathbf{n},\mathbf{m} \rangle} \! S_{\mathbf{n}}^{-}S_{\mathbf{m}}^{+} \!+\! S_{\mathbf{n}}^{-}S_{\mathbf{m}}^{+} \!- \!\sum_{\mathbf{n}} t(S_{\mathbf{n}}^{-}S_{\mathbf{n}}^{+} \!+\! S_{\mathbf{n}}^{z} \!-\!1),
\end{gather}
where \(\mathbf{n}\) and \(\mathbf{m}\) are unit-cell indices, and the left sum runs over nearest neighbors.
The non-ergodic excitation has the same interpretation in this representation as in the 1D case.
Namely, dynamics inside any loop \(\mathcal{S}\) of \(\ket{0_s}\) is isolated from the rest of the lattice.

\section{Conclusions}
\label{sec:conclusion}

In this work, we studied hard-core bosons in the one- and two-dimensional cross-stitch lattices.
The groundstates in these systems admit an analytical construction by first filling the CLS of the single particle flat band and then filling the CLS up to dimers.
The half-filled case corresponds to the Wigner crystal, while higher fillings are described by a mix of Wigner crystal and fully filled dimers.
The crucial property underlying this construction is that the filled CLS acts as an impenetrable barrier for bosons.
As a consequence, any pair of filled CLS in 1D and any closed loop of filled CLS in 2D act as traps for bosons inside, ensuring non-ergodic eigenstates in the spectrum and Hilbert space fragmentation.
We demonstrate similar phenomenology in the diamond chain and its higher dimensional generalizations, as can be straightforwardly verified.
This highlights the unique properties of these models in terms of the connection of the groundstates properties and non-ergodicity.

This raises a question about the properties of other flatband models with interacting bosons.
First, we point out that replacing hard-core bosons with repulsively interacting bosons breaks our construction.
Second, not all flatband models have such nice properties.
For instance, our groundstate construction breaks for the Lieb and kagome lattices, as can be verified explicitly.
It is therefore an interesting open problem: for which classes of flatbands does one find a phenomenology similar to the cross-stitch?
One immediate suggestion is that one needs orthogonal flatbands with CLS occupying two sites.

\begin{acknowledgments}
    AA, SL, SF acknowledge the financial support from the Institute for Basic Science (IBS) in the Republic of Korea through the Project No. IBS-R024-D1.
    We are also grateful to Berislav Bu\v{c}a for multiple useful remarks.
    While preparing this work we became aware of related works~\cite{kwan2023minimal,naik2023quantum}.
\end{acknowledgments}

\appendix

\section{A hard-core boson cannot penetrate a filled CLS}
\label{app:threebosons}

In this Appendix, we demonstrate that a boson cannot penetrate a filled CLS.
For that, we consider a boson placed in unit cell \(0\) next to a filled CLS in unit cell \(ß1\), giving the state \(\ha_0^\dagger\cls\).
We placed the boson on site \(a\) in the unit cell \(0\).
The only term relevant to us in Eq.~\eqref{eq:cs1d} is \(h_0^\dagger\), which moves bosons from the cell \(0\) to the cell \(1\).
All the other terms in the Hamiltonian affect other unit cells and are not relevant for our problem.
Therefore, we compute
\begin{align*}
    h_0^\dagger \ha_0^\dagger\cls &\propto (\ha_1^{\dagger} + \hb_1^{\dagger})(\ha_0 + \hb_0)\ha_0^\dagger(\ha_1^\dagger - \hb_1^\dagger)\vac \\
    &= (\ha_1^{\dagger} + \hb_1^{\dagger})(\ha_1^\dagger - \hb_1^\dagger)\ha_0 \ha_0^\dagger\vac = 0,
\end{align*}
due to destructive interference.
The argument is the same for bosons placed in \(\hb_0\) and also in the 2D model, where the interference happens independently in the space directions.

\section{Mapping of spins onto hard-core bosons}
\label{app:spinToCS}

We provide here the details of the mapping from hard-core bosons to spins.
It is more convenient to work out the mapping from spin to bosons.
We also demonstrate how the Hamiltonian of Ref.~\onlinecite{schecter2019weak} maps onto our problem.
The Hamiltonian \(\mh\) for the spin-1 chain, as given in Ref.~\onlinecite{schecter2019weak}, consists of two components: \(\mh \!=\! \mh_{\mathrm{hop}} \!+\! \mh_{\mathrm{loc}}\), the nearest-neighboring interaction term \(\mh_{\mathrm{hop}}\) and the local field term \(\mh_{\mathrm{loc}}\).
They are defined as
\begin{gather}
    \mh_{\mathrm{hop}} = J\sum_{\langle i,j \rangle}S_{i}^{x}S_{j}^{x} + S_{i}^{y}S_{j}^{y}, \\
    \mh_{\mathrm{loc}} = h\sum_{n\in\Lambda}S_{n}^{z} + D\!\sum_{n\in\Lambda}(S_{n}^{z})^{2}.
\end{gather}
We first inspect the nearest-neighboring interaction term.
Using the definition of the ladder operators~\eqref{eq:ladder_op} and Eq.~\eqref{eq:spinladder}, we recover the inter-cell hopping term of the cross-stitch lattice,
\begin{align}
    \mh_{\mathrm{hop}} &= J\sum_{n\in\Lambda}(S_{n}^{x}S_{n+1}^{x} \!+\! S_{n}^{y}S_{n+1}^{y}) \!+\! (S_{n+1}^{x}S_{n}^{x} \!+\! S_{n+1}^{y}S_{n}^{y}) \notag \\
    &= J\sum_{n\in\Lambda}\left[\vphantom{\sum}(S_{n}^{x}\!-\!iS_{n}^{y})(S_{n+1}^{x}\!+\!iS_{n+1}^{y}) \right. \notag \\
    &\left. \phantom{\sum\sum\sum\sum}{} +  (S_{n+1}^{x}\!-\!iS_{n+1}^{y})(S_{n}^{x}\!+\!iS_{n}^{y})\right] \notag \\
    &= J\sum_{n\in\Lambda}S_{n}^{-}S_{n+1}^{+} + S_{n+1}^{-}S_{n}^{+} \notag \\
    &= J\sum_{n\in\Lambda} \left[\vphantom{\sum} (\hat{a}_{n}^{\dagger} + \hat{b}_{n}^{\dagger})(\hat{a}_{n+1} + \hat{b}_{n+1}) \right. \notag  \\
    &\left. \phantom{\sum\sum\sum\sum}{} + (\hat{a}_{n+1}^{\dagger} + \hat{b}_{n+1}^{\dagger})(\hat{a}_{n} + \hat{b}_{n})\right]
\end{align}
Next, we investigate the local field term.
For spin-1, \(\mathbf{S}\cdot\mathbf{S} = 2I\) holds.
Similarly, we derive the onsite potential and the intra-unit cell hopping term of the cross-stitch lattice using the ladder operators~\eqref{eq:ladder_op} and Eq.~\eqref{eq:spinladder},
\begin{align}
    \mh_{\mathrm{loc}} &= h\sum_{n\in\Lambda}S_{n}^{z} + D\sum_{n\in\Lambda}(2 - S_{n}^{-}S_{n}^{+} - S_{n}^{z})  \\
    &= (h - D)\sum_{n\in\Lambda}\left(1 - (\hat{a}_{n}^{\dagger}\hat{a}_{n} + \hat{b}_{n}^{\dagger}\hat{b}_{n})\right) \notag \\
    &\phantom{\sum\sum\sum\sum}{}+ D\sum_{n\in\Lambda}\left(2 - (\hat{a}_{n}^{\dagger} + \hat{b}_{n}^{\dagger})(\hat{a}_{n} + \hat{b}_{n})\right) \notag \\
    &= \sum_{n\in\Lambda}\left[\vphantom{\sum}(h + D) - D(\hat{a}_{n}^{\dagger}\hat{b}_{n} \!+\! \hat{b}_{n}^{\dagger}\hat{a}_{n}) 
    - h(\hat{a}_{n}^{\dagger}\hat{a}_{n} \!+\! \hat{b}_{n}^{\dagger}\hat{b}_{n})
    \vphantom{\sum}\right] \notag
\end{align}
Choosing \(J = -1\), \(h = 0\), and \(D = -t\), the spin-1 XY chain is reduced to the model derived in Eq.~\eqref{eq:spin1xy} which is expected to have quantum many-body scars.
Moreover, with the consideration of CLSs, we also observe the phenomenon of Hilbert space fragmentation.

\section{1D diamond chain}
\label{app:dia}

\begin{figure}
    \includegraphics[width = \columnwidth]{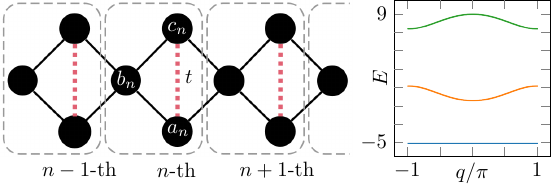}
    \caption{
        (Left) Schematic of the 1D diamond lattice.
        Three sites per unit cell.
        (Right) Energy bands of 1D diamond lattice.
        Here we set \(t = -5\).
    }
    \label{fig:1d_dia_bands}
\end{figure}

Another model displaying properties similar to the cross-stitch chain is the 1D diamond chain, see Fig.~\ref{fig:1d_dia_bands}, with n.n. hoppings and vertical hopping \(t\) (red dashed line on Fig.~\ref{fig:1d_dia_bands}):
\begin{gather}
    \label{eq:dc-ham-hcb}
    \mh = -\sum_{n} \hh_{n} + \hh_{n}^{\dagger} + \hat{t}(\hv_{n} + \hv_{n}^{\dagger}), \\
    \hh_{n} = (\ha_{n}^{\dagger} + \hat{c}_{n}^{\dagger})(\hb_{n} + \hb_{n+1}) \quad\text{and}\quad \hv_{n} = \ha_{n}^{\dagger}\hat{c}_{n}.
\end{gather}
For \(t < -2\) the flatband is the groundstate in the single particle model and the flatband energy, \(E_{\mathrm{FB}} = t\), is gapped away from the other bands.
The flatband is also orthogonal, and different CLS do not overlap.

Now, we briefly summarize the results for the groundstate and its energy.
The construction of the groundstate follows closely the procedure of the one-dimensional cross-stitch lattice for the one-dimensional diamond lattice outlined in Sec~\ref{sec:gs}.
First, CLSs are gradually filled upto filling fraction \(\nu = 1/3\) (since unit cell contains 3 sites in this model).
For \(\nu = 1/3\), we obtain again the Wigner crystal.
For \(1/3 \leq \nu \leq 2/3\), we fill the bottleneck sites with connectivity 4.
This is possible since the filled CLS act as impenetrable barriers, similarly to the cross-stitch case.
Beyond \(\nu=2/3\), CLSs are gradually replaced with fully-filled dimers, just like in the \(\nu>1/2\) case of the cross-stitch.
The groundstate energy \(E_{\mathrm{GS}}\) is determined by the contribution of the CLSs only:
\begin{gather}
    \label{eq:gs_1d_dia}
    E_\mathrm{GS} = 
    \begin{dcases}
        3t\nu N, &\quad \nu \leq 1/3\\
        tN, &\quad 1/3 \leq \nu \leq 2/3\\
        3tN(1-\nu), &\quad 2/3 \leq \nu \leq 1
    \end{dcases}
\end{gather}
This implies macroscopic degeneracies, which follows also from the freedom in filling the CLS, the bottleneck sites, or the dimers in the above construction of the groundstate.

\begin{figure}
    \includegraphics[width = \columnwidth]{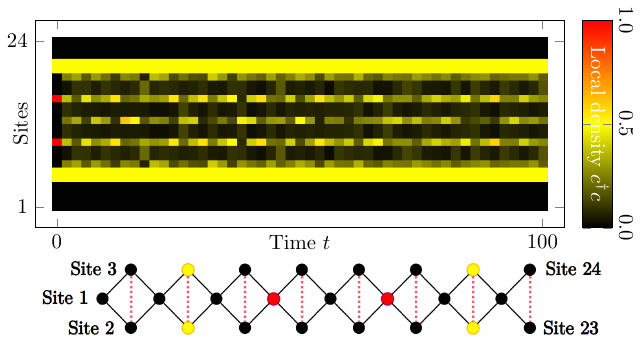}
    \caption{
        The time evolution of the initial wavefunction with four hard-core bosons on the 1D diamond lattice is plotted.
        Here, we set \(t = -5\).
        At time zero, two CLSs are located at the second and the seventh unit cells (yellow) with two hard-core bosons positioned at the 10th and the 16th sites (red).
        The CLSs are fixed as time evolves and two hard-core bosons are strictly confined.
    }
    \label{fig:1d_dia_2hc}
\end{figure}

Since filled CLS act as impenetrable barriers, the diamond chain also possesses non-ergodic excitations like the cross-stitch model.
An example of configuration the leads to caging is shown in Fig.~\ref{fig:1d_dia_2hc}: the two filled CLS marked in yellow trap 2 hard-core bosons placed in between.

Unlike the cross-stitch case, there is no well-defined mapping to spin-1 in the case of the diamond lattice.
Instead, the 1D diamond chain maps onto the spin-1-spin-1/2 (hard-core boson) model,
\begin{align}
    \mh &= -\!\sum_{n} \hat{T}^{-}_{n}(\hb_{n} \!+\! \hb_{n+1}) \!+\! (\hb_{n}^{\dagger} \!+\! \hb_{n+1}^{\dagger})\hat{T}^{+}_{n} \notag \\
    &\phantom{\sum\sum\sum\sum}{}-\!\sum_{n}t\left(\hat{T}^{-}_{n}\hat{T}^{+}_{n} \!+\! \hat{T}^{z}_{n} \!-\! 1\right),
\end{align}
where \(\hat{T}_{n}^{-} \!=\! \ha_{n}^{\dagger}\!+\! \hat{c}_{n}^{\dagger}\) and \(\hat{T}_{n}^{+} \!=\! \ha_{n} \!+\! \hat{c}_{n}\).
The CLSs corresponds to singlet states.
The CLS at the \(n\)~{th} unit cell for the diamond chain is defined as
\begin{gather}
    \cls_{n} = \frac{\hat{a}_{n}^{\dagger} - \hat{c}_{n}^{\dagger}}{\sqrt{2}} \vac.
\end{gather}

\section{1D diamond chain with magnetic flux}
\label{sec:appendix:diaflux}

The diamond chain with no vertical hopping, \(t=0\), also retains a flatband in the presence of a magnetic field.
However, the flatband is no longer orthogonal: eigenstates \(\cls_{n}\) occupy two unit cells and overlap non-trivially:
\begin{gather}
    \mh \!=\! -\!\sum_{n}(\hb_{n}^{\dagger} \!+\! \hb_{n+1}^{\dagger})\ha_{n} \!+\! (\hb_{n}^{\dagger} \!+\! e^{-i\phi}\hb_{n+1}^{\dagger})\hat{c}_{n} \!+\! \mathrm{h.c}., \\
    \cls_{n} \!=\! \frac{1}{\sqrt{4}}\left(\ha_{n}^{\dagger} \!- \hat{c}_{n}^{\dagger}\! + e^{-i\phi}\ha_{n+1}^{\dagger} \!- \hat{c}_{n+1}^{\dagger}\right)\vac.
\end{gather}
The flatband energy \(E_{\mathrm{FB}}\) is precisely zero and no longer corresponds to the groundstate energy.
One can still construct a many-body eigenstate by filling non-overlapping CLS, up to filling fraction \(\nu=1/6\).
The Wigner crystal is obtained for \(\nu=1/6\):
\begin{gather}
    \eig = \prod_{n = 1}^{N/2}\cls_{2n} \hspace{0.5em}\text{or}\hspace{0.5em}\prod_{n = 1}^{N/2}\cls_{2n-1}.
\end{gather}
The next question is whether it is possible to fill up the empty bottleneck sites in CLSs and keep an eigenstate.
The answer is no; doing so does not yield an eigenstate of the Hamiltonian.
This is easily verified by a straightforward calculation starting with the following state:
\begin{gather}
    \kpsi = \prod_{j=1}^{N/2}\kpsi_{2j}\,, \qquad \kpsi_{2j} \!= \hb^{\dagger}_{2j}\cls_{2j-1}.
\end{gather}
Acting with \(\mh\) on \(\kpsi\), we obtain a different state due to non-trivial hopping.
Therefore hard-core bosons are not trapped, as they can escape through the bottleneck sites and there are no non-ergodic excited states.

\bibliography{general,flatband,frustration,mbl,ergodicity,local}

\end{document}